# Conducting A/B Experiments with a Scalable Architecture

Andrew Hornback
Georgia Institute of Technology
ahornback6@gatech.edu

Sungeun An
Georgia Institute of Technology
sungeun.an@gatech.edu

Scott Bunin
Georgia Institute of Technology
sbunin3@gatech.edu

Stephen Buckley
Georgia Institute of Technology
sbuckley@gatech.edu

John Kos
Georgia Institute of Technology
jkos3@gatech.edu

Ashok Goel
Georgia Institute of Technology
ashok.goel@cc.gatech.edu

## ABSTRACT
A/B experiments are commonly used in research to compare the effects of changing one or more variables in two different experimental groups - a control group and a treatment group. While the benefits of using A/B experiments are widely known and accepted, there is less agreement on a principled approach to creating software infrastructure systems to assist in rapidly conducting such experiments. We propose a four-principle approach for developing a software architecture to support A/B experiments that is domain agnostic and can help alleviate some of the resource constraints currently needed to successfully implement these experiments: the software architecture (i) must retain the typical properties of A/B experiments, (ii) capture problem solving activities and outcomes, (iii) allow researchers to understand the behavior and outcomes of participants in the experiment, and (iv) must enable automated analysis. We successfully developed a software system to encapsulate these principles and implement it in a real-world A/B experiment.

## Keywords
Software architectures, Laboratory experiments, Data analytics, AI, A/B experiments, citizen science, conceptual modeling, cognition, education, higher education, human interaction, learning, online learning, virtual assistant, Virtual Experimentation Research Assistant.

## 1. INTRODUCTION
The A/B research methodology allows a researcher to set up a randomized controlled experiment with two groups, A and B, one serving as a control and another as an experiment group, to test two versions of something to determine which performs better. Through changing one or more well-defined characteristics in the experiment group and keeping everything else constant between the two groups, A/B experiments are used to test hypotheses about the effect of those characteristics.

The A/B experimental design has been a staple in academic research for decades, and it recently has seen a surge in popularity in industry as well. As just one example, in 2012, an engineer at Microsoft working on the Bing search engine came up with an idea to change the way advertisement headlines were displayed in search results. After more than six months of the idea remaining dormant given its perceived low priority compared to other ideas that were generated for the problem, the idea was implemented using an online controlled A/B experiment. Within hours, the new method proved its value and was able to generate insights that could increase annual revenues by more than $100 million without hindering the current user experience [1]. Other companies, such as Amazon, Facebook, and Google, have also realized the value of conducting controlled A/B experiments [1]. Each of these companies now conduct more than 10,000 experiments annually online and have the capacity to involve millions of users in each experiment [1]. According to Forbes, companies that are not A/B testing may be lagging direct competitors. Companies can use results from A/B testing to make informative decisions regarding content engagement, reduce the bounce rates of users visiting their websites, and convert more customers to buyers [2].

In academia, A/B experiments are used extensively in research on learning and education. Education can benefit in analogous ways as industry. Results of A/B experiments can help educators understand more effective ways to teach students, understand differences in learning styles, and help students remain engaged in learning. However, A/B experiments can require lots of manual activity, be error prone, and time consuming. One way in which these challenges have been addressed in academia is to try and automate A/B testing. For example, Tamburrelli & Margara framed the A/B testing problem as a search-based software engineering endeavor and concluded that an automated solution for A/B experiments was possible [3]. Recently developed software tools for A/B testing in educational settings, such as UpGrade, were designed to fill voids in other off-the-shelf A/B testing systems related to issues such as the ability to assign students to a group based on criteria such as a particular school class. The engineers who developed UpGrade specifically noted that field testing instructional improvement through A/B testing enabled highly important opportunities for learning engineers to improve learning outcomes at a faster rate than via the traditional educational research cycle [4].

While not all researchers and organizations have the resources that large technology companies such as Google and Microsoft possess, they all can benefit directly from A/B experiments in the same manner. To do so, software architecture that can easily enable A/B experiments has the potential to serve as a mechanism to reduce gaps related to constraints on resources. However, it is not enough to build software that can support learning in the case of academia or consumer analysis. In the case of industry, by enabling A/B experiments to be easily conducted, such systems must have built in components to make analytical insights easily understandable and obtainable by users.

Despite the importance of these goals and the success of A/B experiments, there is only modest research on a generalizable, principled approach to developing software architectures for conducting A/B experiments. There exists a need for such an approach, as it would assist researchers and organizations with creating A/B experiment software quickly. In the paper, "Improving Library User Experience with A/B Testing: Principles and Process" Scott W.H. Young [5] outlines a step-by-step framework for conducting A/B testing, and it includes the following steps:

1) Define a research question,
2) Refine the question with user interviews,
3) Formulate a hypothesis, identify appropriate tools, and define test metrics,
4) Set up and run an experiment,
5) Collect data and analyze results, and
6) Share results and make decision.

We posit that any automated A/B testing software should enable researchers to conduct all these steps, and it should especially aid with steps 4 to 6. As such, we posit that the requirements for building a successful software system that incorporates the entire value chain of A/B experiments should encompass components that endow the following principles:

## 1.1 Retain Typical Properties of A/B Experiments

Software designed to rapidly conduct A/B experiments must retain the ability to test one or more experimental variables, allow the researcher to assign groups manually or randomly, run control and experiment tests simultaneously, and produce measurable, verifiable results.

This principle also ensures that one of the primary voids (mentioned by engineers at Carnegie Learning and PlayPower Labs who developed the aforementioned product UpGrade) in widely available A/B testing software at the time of UpGrade's development is filled. Manual assignment allows instructors to assign learners in an education setting based on class, school, etc., in order to guarantee that students in the same class or other allotment are tested under the same controllable conditions - eliminating a key issue the engineers noted was missing from several systems at the time that can bias results [4].

## 1.2 Capture Problem Solving Activities and Outcomes

Control and experiment groups in A/B testing are designed to test the effect of variables in two settings. The activities that human subjects engage in during the process must be able to be captured for analysis along with the outcomes. Experimental outcomes can lead to acceptance or rejection of the hypothesis being investigated, but if the problem-solving activities incorporated deviate from the experimental design, the outcomes may lack robust support. Databases and other traditional software infrastructure can be designed to fit the data structure needs of an experimental ecosystem to holistically capture data generated by experiment participants that can then be used for analysis.

## 1.3 Must Allow Researchers to Understand Behavior and Outcomes

Interpreting behaviors and outcomes extends the data capturing principle. The data must be captured in a way so that meaning is not lost when conveyed to researchers. Well-designed data schemas for technologies such as databases and modern data structures such as JSON (JavaScript Object Notation) used to store and standardize large amounts of data are essential. Not only do well designed data schemas make the extract, transform, and load (ETL) process used in modern data pipelines more seamless, but can ensure that data collected clearly meets the experiment goals of the researcher.

## 1.4 Must Allow Automated Analysis

Automated data analysis enables researchers to spend more time on the critical parts of the experiment value chain hypothesis generation, experiment design and implementation, and evaluating the results. Thus, a system that has automatic data analysis features in its technology infrastructure is critical in developing software that helps researchers conduct A/B experiments.

We hypothesize that software designed to conduct A/B experiments that harmonizes technological requirements with these four principles can serve as a generalizable, domain agnostic framework in the software development lifecycle. To test this hypothesis, we conducted an A/B experiment using an interactive modeling tool called VERA (Virtual Experimentation Research Assistant; vera.cc.gatech.edu; [6], [7], [8]) developed by the Design & Intelligence Laboratory at the Georgia Institute of Technology and evaluated how well the principles were conveyed.

## 2. VERA

VERA (http://vera.cc.gatech.edu/) is a virtual laboratory for learning about the scientific way of thinking [5], [6], [7].

## 2.1 History of VERA Development

Since 2007, VERA and its previous incarnations ACT [9], EMT [10], and MILA [11], have been used for a variety of learners in multiple settings. Teacher guided middle school science courses, laboratory sections of college undergraduate biology courses, REU summer schools, summer internship programs at museums, adult learners in online programs, citizen scientists seeking to make sense of environmental data, and globally distributed learners of unknown demographics engaged in self-directed learning have all used VERA. In addition, Smithsonian Institute's Encyclopedia of Life (EOL) website (http://www.eol.org) provides direct access to VERA to millions of visitors each year. Lifelong learning is one of the longstanding goals of education [11], [12] and is expounded by VERA through its free and open access to all learners.

Learning about the scientific way of thinking is another long-standing goal of education [14], [15]. VERA is an inquiry-based modeling environment designed to enable learners to explore ecological and other complex systems by performing "what-if" experiments to either explain the behavior of an existing system or attempt to predict the outcome of structural changes to one. Learners use conceptual models and run agent-based simulations of these models to conduct these experiments while engaging in the scientific way of thinking.

The cognitive theory of learning in the VERA project is based on the following constructs:

1) Science makes progress through inquiry-driven model construction, evaluation, and revision [16].
2) Scientific learning is enhanced through engagement with authentic scientific practices [17], [18].
3) Inquiry-driven modeling enhances learning about science [19], [20].

## 2.2 VERA Interface and Operation

Learners in VERA typically begin by identifying atypical or abnormal phenomena such as the overpopulation of a species in an ecosystem and forming one or more hypotheses conjecturing the causes of the observation.

To evaluate a hypothesis, learners develop conceptual models by adding components and relationships among them as shown in Figure 1A. Conceptual models constructed in VERA build on Component-Mechanism-Phenomenon (CMP) models [21] that arose from Structure-Behavior-Function models [22]. Components in VERA serve an analogous structure as a node in a graph and can be either biotic or abiotic with an associated set of variables. Biotic components are defined by basic parameters consisting of lifespan, body mass, starting population, offspring count, reproductive maturity, reproductive interval, and minimum population; as well as advanced parameters consisting of photosynthesis rate, assimilation efficiency, move velocity, respiratory rate, move direction, and carbon biomass. Abiotic substances are defined by the parameters amount, minimum amount, and growth rate. The relationships between components serve an analogous structure as an edge in a graph and are used to describe casual relationships among components in an ecological system using the GloBI ontology of biotic interactions [23]. Examples of such primitive component-component interactions include consumes (one biotic organism consuming another), produces (a biotic organism producing an abiotic substance), and destroys (an abiotic substance harming a biotic organism) as shown in Figure 1B.

Once a user has added components and established relationships among them, parameters can be set as shown in Figure 1C. VERA provides contextualized domain knowledge for ecology via the "Lookup EOL" feature. EOL is the world's largest aggregated and curated database of species data with almost two million species and eleven million attribute records in the biological domain [24].

After constructing a conceptual model, the learner may run agent-based simulations to evaluate the conceptual model as shown in Figure 1D. VERA automatically generates the agent-based simulations using the NetLogo platform directly from the learner's conceptual model (https://ccl.northwestern.edu/netlogo/) [25]. VERA also provides functionality to navigate the simulation and a legend to interpret the graphical data as shown in Figure 1E and Figure 1F respectively.

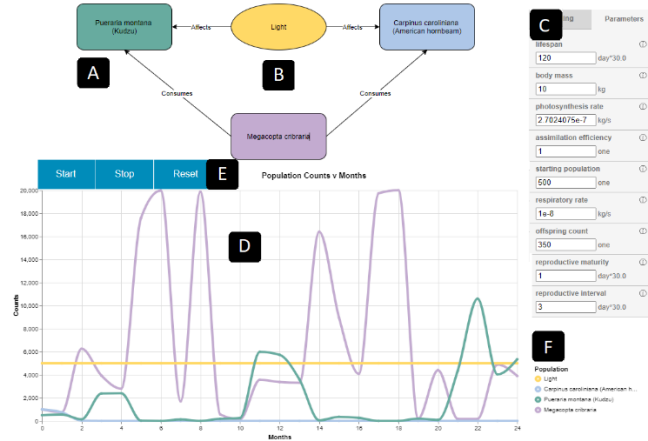

**Figure 1: Operations in VERA primarily consists of model construction, parameter setting, and simulation.**

## 3. DEVELOPING A/B EXPERIMENT INFRASTRUCTURE

To expand VERA from an online laboratory to a tool capable of enabling researchers to conduct rapid A/B experiments with different groups of learners in different learning contexts, we designed and implemented software that met the requirements outlined previously: i) retain typical properties of A/B experiments ii) capture problem solving activities and outcomes, iii) allow researchers to understand behavior and outcomes, and iv) allow automated analysis [26].

The expanded version of VERA features a Researcher mode that encompasses infrastructure that collectively meets the requirements without sacrificing the established user experience that has been part of VERA historically. Model construction, agent-based simulation, hypothesis testing, and all other aspects of VERA did not require direct modifications to implement the new software features designed for A/B experiments.

Researchers begin by creating an experiment. While creating an experiment, they can enable and disable certain features for the two experiment groups. These features include advanced parameters, cloning, exemplar models, lookup EOL, and simulation. Advanced parameters are values associated with model components and include photosynthesis rate, assimilation efficiency, move velocity, respiratory rate, move direction, and carbon biomass. Cloning involves taking a pre-existing model and making a copy to initialize a new model from. Exemplar models are template models available in VERA that are often cloned. Lookup EOL allows users to retrieve parameter values for a biotic component from the Encyclopedia of Life as opposed to initializing the parameters on their own. Finally, simulation allows users to run the models constructed.

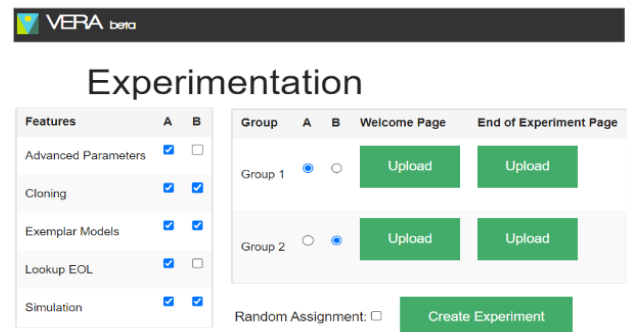

**Figure 2: The VERA Experimentation page allows researchers to quickly design an A/B experiment.**

After choosing which features to enable or disable (the variables to test), a researcher can then upload two PDF files to the system. The first is a welcome page, which can be used to give human subjects background on the experiment, instructions, and other information. The second is an exit page that users see upon completion of the experiment. Next, researchers can choose to enable random assignment for the experiment (as opposed to manual assignment). If researchers do not enable random assignment, they will be given two URLs (one for each group) and assignment is done based on which group's URL was used. If researchers do enable random assignment, one URL is generated, and VERA randomly assigns a user to an experiment group upon going to the URL. The experimental design ensures that the software instills the first principle outlined, retaining the typical properties of an A/B experiment. Randomization has become an established norm in software development [27] but is less commonly observed in educational settings. Thus, the manual setting helps to properly mirror the contexts in which education often takes place, which is in groups such as classrooms [28]. Furthermore, the setting can help to discover whether individual performances of those learning from a common instructor are independent or if group effects exist.

Figure 3: A sample view after a researcher creates an experiment with manual assignment. Custom URLs for each experiment group are generated. Once an experiment subject logs into the URL, the version of VERA he or she accesses will be modified based on the conditions set by the researcher.

After a researcher conducts an experiment, they can download the model data generated by each user that participated. These data contain information such as the structure of the model, including all components and parameters along with the relationships among them, the simulation results, etc. Descriptive statistics such as the total number of learners in each experiment group, the total number of models created in each group, and the total session time are available and have analogous counterparts in other experiments in other domains. Custom metrics such as model complexity (the combined number of components and relationships in a model) along with model variety (the combined number of different components and relationships in a model) are available in VERA.

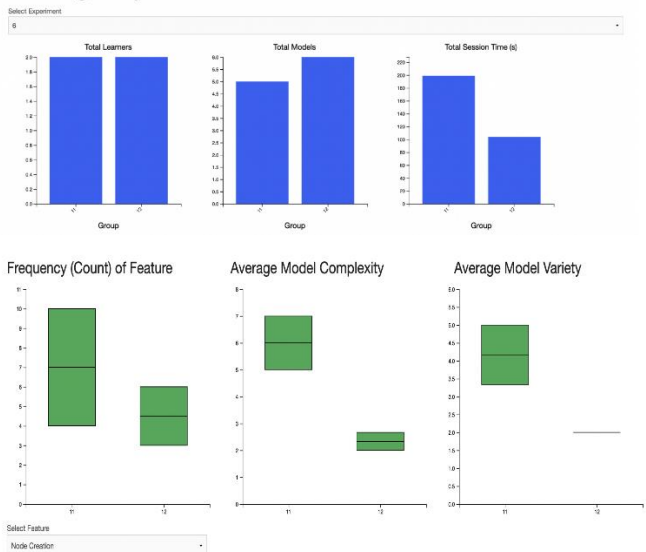

Figure 4: An example of the analytics layer a researcher can see, with total learners, total models, and total session time at the top and frequency, model complexity, and model variety at the bottom.

The analytics layer of the research software implemented in VERA incorporates the remaining three principles described as necessary for building software systems that are designed to rapidly conduct A/B experiments. Problem solving activities and outcomes are stored in a standardized data format in a modern database on the VERA server and are readily available to researchers. Researchers can use the built-in analytics to understand an experiment participant's modeling behavior and outcomes, and the analysis to convey understanding is automated through Python and Java scripts running on the VERA server.

To test the designed architecture and validate not only VERA's ability to conduct A/B experiments, but also to see how well the four outlined requirements holistically fulfill the needs of a software and analytics system to conduct A/B experiments, we designed and conducted an experiment to test the effect of guidance on learning. In this experiment, all features of VERA were left enabled.

## 4. RESEARCH STUDY USING VERA
### 4.1 Setting
Our experiment was conducted at the Georgia Institute of Technology during the spring of 2022. Two classes with very different learning demographics participated in the study:

1) CS 6795: Introduction to Cognitive Science - An online graduate computer science course for adult learners with an age distribution of 24 to 67 years old; and
2) BIOS 4401: Experimental Design and Biostatistics: A blended undergraduate biological sciences course for in-person students, typically 18-24 years old.2.

In the experiment, the variable we set as the experimental variable in the A/B experiment was guided learning. We provided some students with guided instructions on how to use VERA and others without. This allowed us to test the software infrastructure and its ability to meet the four principles by testing a variable external to the software itself, instructor provided guidance, and to also see if we could use the new system to validate previous work on VERA that showed learners who received minimal guidance in an experimental setting implemented more exploratory problem-solving strategies than those who received more guidance [29].

### 4.2 Sampling and Condition Assignments
In both courses, students were separated into two groups and evaluated over two sessions that were at minimum 30 days apart. Students in CS 6795 were placed in teams of up to four students while students in BIOS 4401 worked individually or in teams.

*4.2.1 Phase I in CS 6795, Group A (Unguided).*
Teams of students in Group A were given an ecological system to model using VERA based on an exemplar template model and asked to iteratively build and evaluate the model through agent-based simulation. The model consisted of four components - kudzu, a type of plant that grows on trees and buildings; american hornbeam, a type of tree, kudzu bug, a bug that feeds on kudzu and american hornbeam, and light which has an effect on both kudzu and american hornbeam.

After running multiple simulations, teams downloaded the aggregate simulation data using the new software infrastructure and discussed how the kudzu population shifts in relation to the other components of the model and whether the simulation data was normally distributed. Based on this analytical discussion, teams generated a list of hypotheses to test to shift the peak of the kudzu population right on the distribution curve in order to have a higher population on average.

For each hypothesis, teams then modified the model by changing parameter values and model components as necessary to test the validity of each hypothesis. Multiple simulations were performed followed by an aggregation of the simulation data. Based on analysis of the aggregated simulation data, teams accepted or rejected each proposed hypothesis.

### 4.2.2 Phase I in CS 6795, Group B (Guided).
Teams of students in Group B performed the same experiment as Group A, with the difference being that Group B was given explicit instructions for setting simulation parameter values as well as specific hypotheses to tests. The instructions for Group B were provided at the beginning of each session.

Specifically, the three hypotheses were:

1) Decreasing the reproduction rate of kudzu will shift the kudzu distribution peak to the right;
2) Increasing the initial population of the kudzu bug will shift the kudzu distribution peak to the right; and
3) Decreasing the consumption rate of kudzu and the kudzu bug will shift the kudzu distribution to the right.

Providing hypotheses to test with specific parameters to alter for each hypothesis restricted the parameter space in Group B compared to Group A which was free to explore the entire space.

### 4.2.3 Phase II in CS 6795, Both Groups A (Unguided) and B (Guided).
In the second session, transfer learning was evaluated by giving teams a more complex ecological modeling problem scenario using the same model as in Phase I. Group A and Group B were given the same set of instructions.

Teams were first asked to make a new copy of the kudzu model, perform multiple repeated simulations, and record the aggregate simulation data in the same manner as Phase I. Next, teams would analyze how the populations of each of the kudzu, kudzu bug, and american hornbeam changed over time (ie, the kudzu population decreased after 3 months, the kudzu bug population increases for 5 months before beginning to decrease, and the american hornbeam population peaks at month 4). Based on this analysis, teams formed one hypothesis for each team member and recorded a description of the hypothesis along with the independent variable(s) and dependent variable(s) being tested.

To test the newly formed hypotheses, teams selected multiple levels for each independent variable(s), decided how many simulations to run, formed a prediction for the hypothesis result, and finally tested each hypothesis.

Near and far transfer learning were to be evaluated in Phase II based on the parameter choices made. If the experimental design were successful in demonstrating VERA's potential as a tool to evaluate learning, we would expect to see near transfer learning in the guided group, Group B, and far transfer learning in the unguided group, Group A, based on how the parameter choices in Phase I transferred to Phase II.

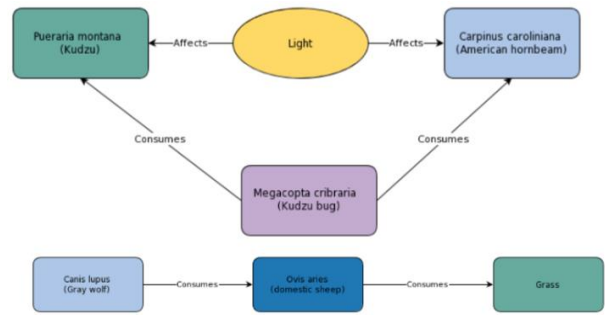

**Figure 5: Ecological system used in Phases I and II of CS6795 (left) and BIOS 4401 (right).**

### 4.2.4 Phase I in BIOS 4401.
Students and teams of students were given a simple ecological system to model using VERA based on an exemplar template model and asked to iteratively build and evaluate the model through agent-based simulation in the same manner as teams of students in CS 6795. The model used was a model of a predator (wolf) and prey (sheep) ecological system that also consisted of a grass component that was consumed by the prey.

Next, each student or team ran 10 simulations without making any changes to the model. The aggregate data from the simulations was then downloaded and the population size of the sheep analyzed. Based on the analysis, each student and team came up with a list of hypotheses that would result in the population distribution peak of the sheep shifting to the right on the distribution curve.

To test their hypotheses, students and teams were free to change any components or parameters needed, similar to Group A, the unguided group in the experiment conducted in CS 6795. To conclude Phase I, students and teams ran simulations to test their hypotheses and recorded the results.

### 4.2.5 Phase II in BIOS 4401.
In Phase II, students and teams followed the same instructions in Phase I with the exception that each tested hypothesis involved testing multiple levels, such as low and high, for the independent variable(s) involved.

Based on the experimental design, we considered near transfer learning to be established when a learner could address new problems very similar to those addressed previously, by changing the same component parameter value for example. In contrast, we considered far transfer learning to be established when facing new problems that although were in the same domain, were dissimilar to those previously addressed but were addressed using previous problem solving mechanisms learned in past problems.

As noted, given the experimental design, we would expect to see near transfer learning in Group B of CS 6795 given the guidance provided for hypothesis testing. Contrarily, we would expect far transfer learning to occur in Group A of CS 6795 and BIOS 4401 given the lack of guidance for hypothesis testing. Specifically, the overall parameter space explored and the frequency of parameters changed would be expected to have predictable relationships in Phase II of both classes based on experiences in Phase I. We would use the data capture and analysis tools in the newly designed software infrastructure to assist with this analysis.

## 4.3 Experiments Summary

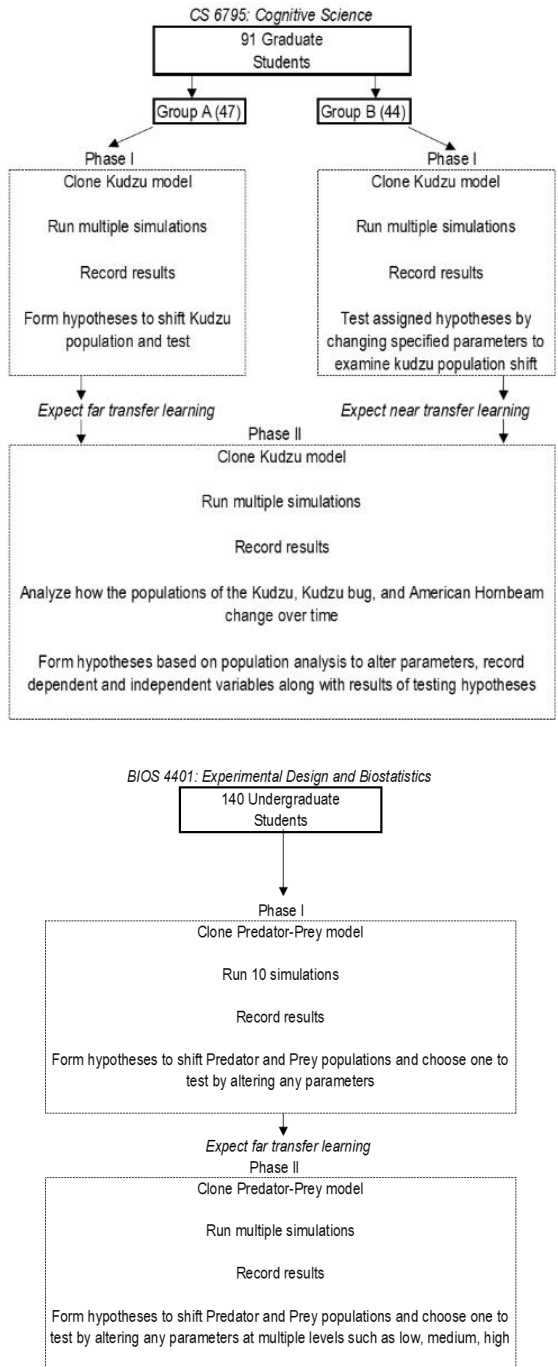

Figure 6: Experiment designs implemented to conduct A/B testing in VERA.

In CS 6795, 91 students comprised the initial experiment population while BIOS 4401 began with 140 students. To examine transfer learning, we examined data produced by each student that was captured in VERA's database layer of the software architecture. The format of the data allowed us to analyze what components and parameter values were used in simulations. The data storage, format, and retrieval process aligned with two of the four principles outlined - capture problem solving activities and allow researchers to understand behavior and outcomes.

## 5. RESULTS

In both BIOS 4401 and CS 6795, our hypothesis about within-domain far transfer learning among students who received minimal guidance and near transfer learning among guided students was evident. Simultaneously, we were able to validate VERA's ability to conduct A/B testing and incorporate the four principles established at the onset of the research software development process.

### 5.1 Parameter Space Analysis

We define parameter space as the total number of unique combinations of model components (such as kudzu in the CS 6795 model and wolf in the BIOS 4401 model) and parameters (such as lifespan) changed in the combined phases for both classes. The total size of the parameter space for BIOS 4401 was 14 and for CS6795 was 35. Parameter space data was downloadable for each user that participated in the experiment, illuminating the second principle of the design architecture, capturing problem solving activities and outcomes.

Table 1: Parameter space for BIOS 4401

| BIOS 4401 - Parameter Space | |
|---|---|
| Canis lupus - lifespan | Grass - lifespan |
| Canis lupus - offspring count | Grass - minimum population |
| Canis lupus - reproductive interval | Grass - starting population |
| Canis lupus - reproductive maturity | Ovis aries - body mass |
| Canis lupus - starting population | Ovis aries - lifespan |
| Consumes - consumption rate | Ovis aries - offspring count |
| Grass - body mass | Ovis aries - starting population |

Unguided students in BIOS 4401 and Group A of CS 6795 explored a much larger portion of the parameter space in both phases of the experiment than the guided students in Group B of CS 6795. In Phase I of BIOS 4401, 78.57% of the parameter space was explored, followed by 71.43% in Phase II. In Group A of CS 6795, 64.52% of the parameter space was explored in Phase I, followed by 66.67% in Phase II. Thus, the two unguided groups explored a large and similar amount of the parameter space. Without being given specific instructions in either scenario to test respective population shifts, far transfer learning is observed given that a large exploration of the parameter space to understand how the conceptual model of the system works was implemented as a learning strategy in both phases.

In contrast, Group B of CS 6795 explored 32.36% of the parameter space in Phase I, followed by 60.0% in Phase II. While the increase in overall parameter space explored was much greater than the unguided groups, Group B still focused heavily on parameters related to reproduction, population, and consumption - as they were instructed to do in Phase I. In Phase I, Group B chose among consumption rate, initial population, minimum population, offspring count, reproductive interval, and reproductive maturity as the component parameter changed 70.00% of the time. In Phase II, this number was 88.89% - indicating that the group used the same underlying processes to solve the similar problem in Phase II as Phase I without introducing a lot of novelty in the problem solving context.

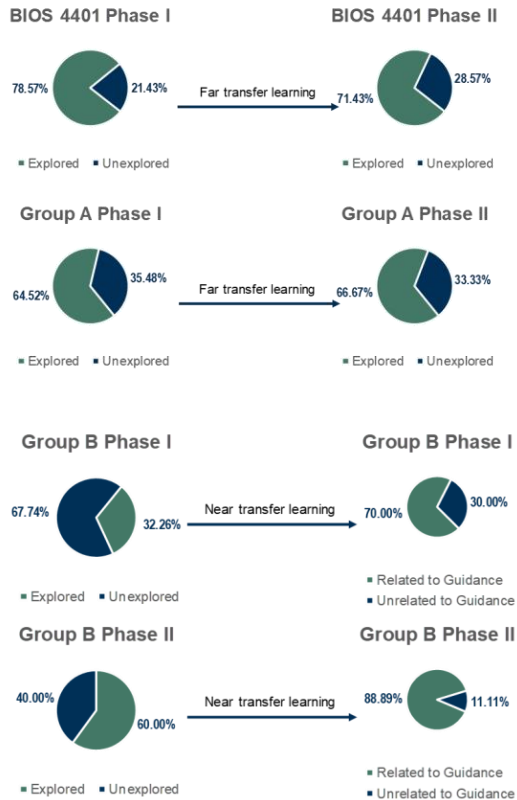

**Figure 7: CS 6795 Group A and BIOS 4401 showed far transfer learning - exploring a large portion of the parameter space as a learning strategy in both phases. CS 6795 Group B showed near transfer learning - even with an increase in the parameter space explored, learners still focused on parameters related to guidance from Phase I.**

## 6. DISCUSSION

The guided vs unguided experiment conducted provided insight into the performance of the newly designed software infrastructure as a data capture and analysis tool for assisting researchers conduct A/B experiments. In addition to helping validate that the software fulfills the requirements of the four principles outlined as baseline requirements for an analytics-driven software infrastructure for conducting rapid A/B experiments, the experiment also validated earlier work conducted on VERA that showed learners receiving minimal guidance showed more exploratory behavior [28].

In addition to establishing the four principles outlined, for future work, we hope to engage with researchers and instructors who use VERA to enable various stakeholders to have a voice in the construction of analytics features. We are currently working with instructors at North Georgia Technical College (NGTC) to create a version of the researcher mode specifically for course instructors. The objective is to provide instructors with a transparent interface in VERA where they can further integrate student data and was kindled with the direct input of the instructors at NGTC after they implemented VERA in a fish and wildlife course.

Also, we are integrating predictive analytics tools that are based on the analysis of all the current models stored in VERA's database. Certain patterns, such as model construction followed by simulation followed by parameter alterations, are typical in VERA. Other patterns, such as hitting the simulation button three times in a row followed by resetting the simulation, can be indicative of a learner not fully understanding how to use the system. Using instructor and researcher input from those who have successfully used VERA in courses or published academic papers, we have developed analytics tools that can cluster modeling behaviors so that we can potentially identify those who may benefit from immediate feedback while using the system.

The predictive analytics tools are being built by mapping sequence behavior patterns using Markov chains. We have discovered fifteen modeling behavior patterns in VERA and assigned these behaviors to three groups: 1. Observation, 2. Construction, and 3. Exploration. While the Observation pattern was most common, models classified as Exploration were of higher quality [30].

Observation behaviors involve starting a model, simulating it, and perhaps altering the parameters. Construction patterns extend Observation patterns by adding components and constructing behavior. Exploration is the pattern in which we ultimately hope to guide all learners to. The Exploration pattern involves starting, constructing, simulating, changing parameters, possibly using EOL, etc., in a cyclical manner that shows full understanding of the modeling process. By analyzing model behaviors of users over time, we will be able to provide feedback to users who have not reached the Exploration stage in areas in which they need additional guidance.

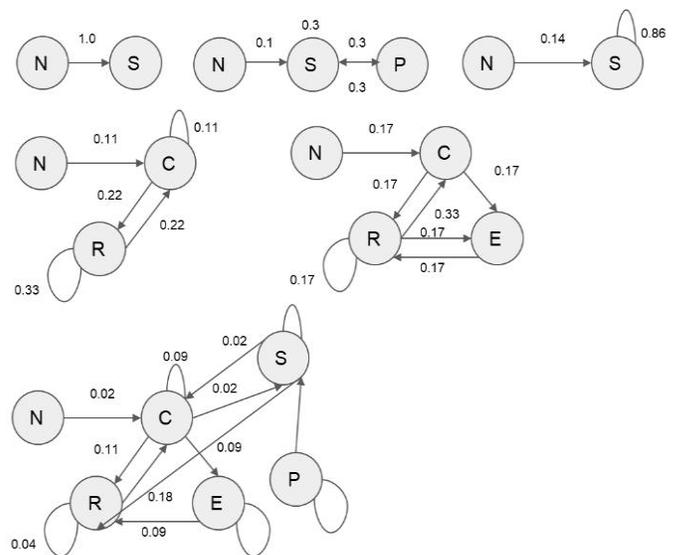

**Figure 8: Modeling behavior patterns found using Markov chains. The top row shows three examples of the Observation behavior, which focuses on simulating a model and perhaps changing parameters. The middle row shows two examples of the Construction behavior, which extends the Observation behavior to include constructing models, revising, etc. The bottom row shows the Exploration behavior which encompasses all actions possible in VERA along with logical transition states among the actions. All numbers represent state transition probabilities between actions. (Legend: N = New (Start), S = Simulation, P = Parameter Change, C = Construction, R = Revision (adding/deleting components, changing relationships), E = EOL)**

While some of the future goals related to transparency and feedback in analytics are still in development, the current system does reliably and transparently model student competencies, as shown in the A/B experiment conducted. The ability to capture and understand learning behaviors was apparent, and we will continue to build upon the current architecture with the goal of including as large a variety of stakeholders as feasible.

## 7. CONCLUSION

Conducting A/B experiments is an effective research method in academia and industry. When building technologies ranging from those that support learning such as VERA in an academic setting to those that are used to make search engine advertisement decisions in an industry setting such as Microsoft, incorporating the ability to build technologies that easily enable A/B experiments without hindering the intended user experience is vital for deriving analytical insights that can be used in decision making.

In developing software infrastructure for VERA to conduct and support A/B experiments, we developed technology components that enabled us to successfully accomplish four principles that can serve as at minimum baseline guiding principles for other such systems:

1) Retain the typical properties of A/B experiments.
2) Capture problem solving activities and outcomes.
3) Allow researchers to understand behavior and outcomes.
4) Allow automated analysis.

We conducted an A/B experiment using different types of learners, online and in-person, undergraduate and graduate, comparing the effects of guidance as the experimental condition. The data collection and automatic analysis provided by the software additions to VERA provided capabilities to analyze how human subjects in each condition solved the assigned problems. The behavior and outcomes of the human subjects were observable by using descriptive analytics that can be analogously implemented in other systems paired with novel statistics such as model complexity and model variety that enable researchers to understand more about how model construction and simulation relates to whether students in given conditions were successful in implementing learning strategies that led to accepting a hypothesis.

An important feature of our work is that it is generalizable to other domains and software architectures. Researchers who can create efficient infrastructure for capturing and automatically analyzing applicable data for their area can focus more time on the high value-added elements of the experiment value chain and scale their research quickly. Particularly, educators can focus more on how specific students differ from each other and what will better their education. Prior to implementing the current research features in VERA, analyzing the data would require a lot of manual effort that would have hindered progress and potentially be prone to errors.

## 8. ACKNOWLEDGMENTS

This research was supported in part by an US NSF grant #1636848 and now is partly supported by an US NSF National AI Institutes grants #2112532 and #2247790 (AI-ALOE: National AI Research Institute for Adult Learning and Online Education.) We thank Spencer Rugaber for his contributions to the VERA project.

## 9. REFERENCES

[1] R. Kohavi and S. Thomke, "The Surprising Power of Online Experiments," Harvard Business Review, 16-Sep-2020. [Online]. Available: https://hbr.org/2017/09/the-surprising-power-of-online-experiments. [Accessed: 26-Sep-2022].

[2] J. Simpson, "A/B Testing: The Benefits And How To Use It Efficiently," Forbes, 12-Mar-2020. [Online]. Available: https://www.forbes.com/sites/forbesagencycouncil/2020/03/12/ab-testing-the-benefits-and-how-to-use-it-efficiently/?sh=49de0e7786d4. [Accessed: 22-Sep-2022].

[3] Tamburrelli, Giordano & Margara, Alessandro. (2014). Towards Automated A/B Testing. 10.1007/978-3-319-09940-8_13.

[4] Ritter, Steven & Murphy, April & Fancsali, Stephen & Fitkariwala, Vivek & Patel, Nirmal & Lomas, Derek. (2020). UpGrade: An Open Source Tool to Support A/B Testing in Educational Software

[5] N Young, S. W. (2014). Improving library user experience with a/B testing: Principles and process. Weave: Journal of Library User Experience, 1(1). https://doi.org/10.3998/weave.12535642.0001.101

[6] An S., Bates R., Hammock J., Rugaber S., Goel A. (2018) VERA: Popularizing Science Through AI. In: Penstein Rosé C. et al. (eds) Artificial Intelligence in Education. AIED 2018. Lecture Notes in Computer Science, vol 10948. Springer.

[7] An, S., Bates, R., Hammock J., Rugaber, S., Weigel, E., & Goel, A. (2020) Scientific Modeling Using Large Scale Knowledge. In Procs. 21st International Conference on AI in Education (AIED 2020), pp. 20-24.

[8] Goel, A. AI-Powered Learning: Making Education Accessible, Affordable, and Achievable. 2020. Available online: https://arxiv.org/abs/2006.01908.

[9] Vattam, S., Goel, A., Rugaber, S., Hmelo-Silver, C., Jordan, R., Gray, S., & Sinha, S. (2011). Understanding Complex Natural Systems by Articulating Structure-Behavior-Function Models. Educational Technology & Society, 14(1), 66-81.

[10] Joyner, D., Goel, A., Rugaber, S., Hmelo-Silver, C., & Jordan, R. (2011). Evolution of an Integrated Technology for Supporting Learning about Complex Systems: Looking Back, Looking Ahead. In Proc. 11th International Conference on Advanced Learning Technologies, Athens, GA.

[11] Joyner, D., and Goel, A. 2015. Improving Inquiry-Driven Modeling in Science Education Through Interaction with Intelligent Tutoring Agents. In Procs. 20th ACM Conference on Intelligent User Interfaces, 5-16. New York: ACM

[12] Field, J. (2000) Lifelong learning and the new educational order. Trentham Books, UK.

[13] Fischer, G. (1999) Lifelong learning: Changing mindsets. In Proceedings 7th International Conference on Computers in Education, Chiba, Japan, pp. 21-30.

[14] Kuhn, D., Amsel, E., & O'Laughlin, M. (1988). The development of scientific thinking skills. Orlando, FL: Academic Press.

[15] Lehrer, R., & Schauble, L. (2015). The development of scientific thinking. In L. Liben & U. Mueller (Vol. eds.) & R.Lerner (Series ed.), Handbook of child psychology and

developmental science, Vol. 2: Cognitive process. (7th Edition). Hoboken, NJ: Wiley.

[16] Darden, L. (1998). Anomaly-driven theory redesign: computational philosophy of science experiments. In The Digital Phoenix: How Computers Are Changing Philosophy (Bynum, T., & Moor, J., Eds.), pp. 62–78. New York: Blackwell

[17] Clement, J. (2008). Creative model construction in scientists and students. Springer.

[18] Edelson, D. C. (1998). Realising authentic science learning through the adaptation of science practice. In B. J. Fraser & K. Tobin (Eds.), International Handbook of Science Education (pp. 3 17-33 1). Dordrecht, The Netherlands: Kluwer.

[19] VanLehn, K., 2013. Model construction as a learning activity: A design space and review. Interactive Learning Environments 21, 371–413.

[20] White, B., Frederiksen, J., 1990. Causal model progressions as a foundation for intelligent learning environments. Artificial intelligence 42, 99–157.

[21] Joyner, D., Goel, A., Rugaber, S., Hmelo-Silver, C., & Jordan, R. (2011). Evolution of an Integrated Technology for Supporting Learning about Complex Systems: Looking Back, Looking Ahead. In Proc. 11th International Conference on Advanced Learning Technologies, Athens, GA.

[22] Goel, A., Rugaber, S., Vattam, S., 2009. Structure, behavior, and function of complex systems: The structure, behavior, and function modeling language. Ai Edam 23, 23–35.

[23] Poelen, J., Simons, J., Mungall, C., (2014) Global biotic interactions: An open infrastructure to share and analyze species-interaction datasets. Ecological Informatics 24, 148–159.

[24] Parr, C., Schulz, K., Hammock, J., Wilson, N., Leary, P., Rice, J., Corrigan Jr, R.. (2016) Traitbank: Practical semantics for organism attribute data. Semantic Web 7, 577–588.

[25] Wilensky, U., Resnick, M., 1999. Thinking in levels: A dynamic systems approach to making sense of the world. Journal of Science Education and technology 8, 3–19.

[26] Hornback, A., Buckley, S., Kos, J., Bunin, S., An, S., Joyner, D., and Goel, A. A Scalable Architecture for Conducting A/B Experiments in Educational Settings. In Proceedings of the Tenth ACM Conference on Learning @ Scale (L@S '23). Association for Computing Machinery, New York, NY, USA, 373–377. https://doi.org/10.1145/3573051.3596190

[27] Stefan Thomke. 2020. Experimentation works: The surprising power of business experiments. Harvard Business Review Press.

[28] Steven Ritter, April Murphy, and Stephen E. Fancsali. 2020. Managing group random assignment in UpGrade. Submitted to Proceedings of the First Workshop on Educational A/B Testing at Scale (at Learning @ Scale 2020).

[29] An, S., Weigel, E., & Goel, A. (2022, June). Effects of Guidance on Learning about Ill-defined Problems. In International Conference on Intelligent Tutoring Systems. Springer, Cham.

[30] An, S., Rugaber, S., Hammock, J., Goel, A.K. (2022). Understanding Self-Directed Learning with Sequential Pattern Mining. In: Rodrigo, M.M., Matsuda, N., Cristea, A.I., Dimitrova, V. (eds) Artificial Intelligence in Education. Posters and Late Breaking Results, Workshops and Tutorials, Industry and Innovation Tracks, Practitioners' and Doctoral Consortium. AIED 2022. Lecture Notes in Computer Science, vol 13356. Springer, Cham. https://doi.org/10.1007/978-3-031-11647-6_102